\documentclass[floats,floatfix,showpacs,preprintnumbers,amssymb,prd,twocolumn,superscriptaddress,nofootinbib,nolongbibliography,reprint]{revtex4-1}

\usepackage{amssymb,amsmath,verbatim,mathtools,needspace,enumitem,etoolbox,graphicx,physics,microtype,afterpage,xspace,tabularx,lmodern,multirow}
\usepackage{gensymb}
\usepackage{appendix}
\usepackage{tensor}
\usepackage[normalem]{ulem}
\usepackage[dvipsnames, usenames]{xcolor}
\usepackage{xr-hyper}
\definecolor{linkcolor}{rgb}{0.0,0.3,0.5}
\usepackage[unicode, colorlinks=true, linkcolor=linkcolor, citecolor=linkcolor, filecolor=linkcolor, urlcolor=linkcolor, linktocpage, breaklinks]{hyperref}
\usepackage[all]{hypcap}
\usepackage[T1]{fontenc}
\usepackage[utf8]{inputenc}
\usepackage[usenames,dvipsnames]{xcolor}
\hypersetup{colorlinks=true,citecolor=magenta,linkcolor=violet,urlcolor=magenta}

\definecolor{romared}{RGB}{142,0,28}

\newcommand{\be}{\begin{equation}}
\newcommand{\ee}{\end{equation}}

\def\be{\begin{equation}}
\def\ee{\end{equation}}
\newcommand{\beq}{\begin{eqnarray}}
\newcommand{\eeq}{\end{eqnarray}}

\usepackage{makecell}
\usepackage{soul}

\newcolumntype{Y}{>{\centering\arraybackslash}X}

\makeatletter
\newcommand*{\addFileDependency}[1]{
  \typeout{(#1)}
  \@addtofilelist{#1}
  \IfFileExists{#1}{}{\typeout{No file #1.}}
}
\makeatother

\newcommand*{\myexternaldocument}[1]{%
    \externaldocument{#1}%
    \addFileDependency{#1.tex}%
    \addFileDependency{#1.aux}%
}

\myexternaldocument{Supp}

\begin{document}
\title{Magnetic Black Holes: from Thomson Dipoles to the Penrose Process and Cosmic Censorship}

\author{Conor Dyson}
\author{David Pere\~niguez} 
\affiliation{Niels Bohr International Academy, Niels Bohr Institute, Blegdamsvej 17, 2100 Copenhagen, Denmark}

\date{\today}

\begin{abstract}

We consider accretion of charged test matter by rotating, magnetic black holes and discuss a number of aspects in which the interaction of the angular momentum contained in the electromagnetic field and the spin of the hole plays a fundamental role. First, we argue that such a black hole tends to lose its angular momentum by accreting charges while remaining globally neutral. Then, we show that accretion can happen in a superradiant manner via an enhanced Penrose process. In particular, we find that the regions from which energy and angular momentum can be extracted contain the axis of rotation and, in some cases, consist of floating bubbles disconnected from the black hole itself. Finally, we address the question of whether extremal dyonic rotating black holes can be overcharged or overspun via accretion of arbitrary matter, and prove that this can not happen if the null-energy condition holds. We conclude by discussing some future research directions.

\end{abstract}


\maketitle

\section{Introduction} \label{Intro}

Understanding the strong field regime of gravitation, as well as its interaction with matter and other fundamental forces raises as one of the main challenges of modern physics. The complexity of this problem is in sharp contrast with one of the deepest predictions of General Relativity (GR) which establishes that all quiescent black holes in the Universe are uniquely described by a simple solution of the field equations. If one considers the coupling of gravitation to electromagnetism, described by the Einstein--Maxwell theory, then this solution, the Kerr--Newman (KN) black hole \cite{Kerr:1963ud,NPSol}, has four independent parameters consisting of its mass $M$, angular momentum $J$, electric and magnetic charges $Q$ and $P$~\cite{Carter:1971zc,PhysRevLett.34.905,Ruffini:1971bza,Hawking:1973uf}. Given the unquestionable importance of this result, it comes as no surprise that KN black holes have captured the interest of theoretical astrophysicists for decades~\cite{Zerilli:1974ai,Johnston:1974vf,PhysRevD.10.1057,Gerlach:1979rw,Gerlach:1980tx,Chandrasekhar1979OnTM}, and that nowadays in the dawn of gravitational wave astronomy they play a prominent role in searches of beyond vacuum GR physics from ringdown analysis \cite{Carullo:2021oxn,Dias:2021yju}, in modelling signatures of dark matter \cite{DeRujula:1989fe,Perl:1997nd,Holdom:1985ag,Sigurdson:2004zp,Davidson:2000hf,McDermott:2010pa,Cardoso:2016olt,Khalil:2018aaj,Bai:2019zcd,Gupta:2021rod,Kritos:2021nsf}, as well as in providing a well defined set up for non-vacuum numerical simulations of black hole coalescences \cite{Zilhao:2012gp,Zilhao:2013nda,Liebling:2016orx,Bozzola:2020mjx}. Besides, there are reasons to believe that charged black holes could play an important role in high-energy astrophysical phenomena such as cosmic rays \cite{Tursunov:2019oiq}. 

The vast majority of black holes in the Universe are expected to be neutral. On the one hand, Schwinger pair creation and friction with interstellar medium would most likely lead a black hole towards quickly losing any amount of electric charge it may have accumulated \cite{Gibbons:1975kk} and, even though there are well known astrophysical mechanisms through which black holes accrete and retain a net amount of electric charge, this is very small in realistic scenarios \cite{Wald:1974np,Beskin:2000qe,Palenzuela:2011es}. On the other hand, magnetic charges are to date only theoretical (yet robust) predictions. However, there are compelling reasons to believe that magnetic monopoles were produced in the early Universe (primordial monopoles)\cite{Preskill:1984gd}. It is possible that black holes formed at that time could have absorbed some net amount of magnetic monopoles, or that were formed directly from the collapse of the latter, thus turning the magnetic monopole problem of cosmology into a problem of magnetic primordial black holes \cite{Gibbons:1990um}. Since magnetic monopoles are less likely to pair create than electric charges, those black holes would have Hawking-evaporated until reaching extremality and could have remained until today, perhaps contributing to a fraction of the dark matter in the Universe \cite{Stojkovic:2004hz,Bai:2019zcd,Maldacena:2020skw,Kritos:2021nsf}. Besides, from a theorist's perspective it is desirable to retain full generality when possible, so in this work we will allow black holes to rotate and carry electric and magnetic charges without further ado.

From electric-magnetic duality it follows that, in isolation, one can restrict to purely electric black holes without loss of generality. However, if there are more charges the purely electric set up is no more the most general one, e.g.~it does not include, via duality, the interaction of an electric particle and a magnetic black hole. As is well known, the electromagnetic field created by electric and magnetic charges placed together exhibits some remarkable properties, and coupling those to a strong gravitational field is a very interesting problem both from a fundamental and an observational perspective. 

The purpose of this work is to study the interaction of charged test matter with magnetic (or more generally, dyonic) black holes. In Sections \ref{M} and \ref{AC} we revisit the motion of charged particles in the background of a dyonic KN black hole and show that, when immersed in an ionised medium, a magnetic rotating black hole tends to lose its angular momentum by accreting electric charges while remaining globally neutral (unlike the case of magnetised black holes \cite{Wald:1974np}). Together with this paper we have also made available a \texttt{Mathematica} package implementing our solutions for generic plunging and bound trajectories, following the methods in \cite{LabanMutrie, Fujita:2009bp, vandeMeent:2019cam, Gralla:2019ceu, Dyson:2023fws}. In Section \ref{PP} we show that accretion can lead to superradiant phenomena, and that the Penrose process is greatly enhanced if the black hole rotates and possesses magnetic charge. In particular, we show that it is possible to extract both energy and angular momentum from the hole in axisymmetric processes, such as matter ejection along the axis of rotation. We also identify new ``floating'' regions of spacetime from which energy can be extracted (regions where ``negative energy states'' exist) that are disconnected from the black hole and the mechanical ergoregion. Finally, in Section \ref{CC} we consider whether dyonic, extremal rotating black holes could develop a naked singularity by overcharging or overspinning them via accretion, thus incurring into a contradiction with the weak cosmic censorship conjecture. We begin by revisiting the case of in-falling particle matter (first considered in \cite{HGed}), and show it is a spin-spin repulsion mechanism that prevents a violation of cosmic censorship. Then, incorporating recent developments in dealing with magnetic charges in covariant phase-space \cite{Ortin:2022uxa}, we are able to provide a simple proof that, regardless of the nature of the in-falling matter, extremal dyonic black holes cannot be overcharged or overspun as long as the null-energy condition is satisfied. We conclude in Section \ref{D} by discussing our results as well as some future directions.

\section{Dyonic Black Holes and the Motion of Charged Particles} \label{M}

The motion of a point particle with mass $m$, electric charge $e$ and magnetic charge $g$ is governed by the equation
\begin{equation}\label{GeodEq}
u^{a}\nabla_{a}u_{b}=\frac{1}{m}\left(e F_{b a}-g \star F_{b a}\right)u^{a}\, .
\end{equation}
Carter decoupled and solved formally these equations on the dyonic Kerr--Newman spacetime in \cite{Carter:1973rla}.\footnote{This problem has been reconsidered independently later on in the literature, see e.g.~\cite{Grunau:2010gd,Hackmann:2013pva,Russo:2020lah}.} In this section we give an alternative derivation in terms of variables that make manifest gauge- and duality-invariance, are not specific of the dyonic Kerr--Newman solution and have a close relation to the laws of black hole mechanics (see Section \ref{CC} and \cite{Ortin:2015hya,Elgood:2020svt,Ortin:2022uxa}). However, it is convenient to first introduce the dyonic KN solution and some elements of notation.

In Boyer--Lindquist coordinates, the line element takes the familiar form
\begin{equation}
\begin{aligned}\label{KN}
&ds^{2}=-\frac{\Delta-{a^2}\sin^{2}\theta}{\Sigma}dt^{2}-2a\sin^{2}\theta\left(\frac{r^2+a^2-\Delta}{\Sigma}\right)dtd\phi\\
&+\left(\frac{(r^2+a^2)^2 -\Delta a^2\sin^{2}\theta}{\Sigma}\right)\sin^{2}\theta d\phi^{2}+\frac{\Sigma}{\Delta}dr^{2}+\Sigma d\theta^{2}\, , 
\end{aligned}
\end{equation}
where the angular coordinate is canonically normalised, $\phi\sim\phi+2\pi$, and
\begin{equation}
\Delta=r^{2}-2Mr+a^{2}+Q^{2}+P^{2}, \ \ \ \Sigma=r^{2}+a^{2}\cos^{2}\theta\, ,
\end{equation}
while the Maxwell potential can be expressed as
\begin{equation}\label{MaxVec}
    A=-\frac{Q r}{\Sigma}\left(dt-a \sin^{2}\theta d\phi\right)+\frac{P\cos\theta}{\Sigma} \left(a dt-(r^2+a^2)d\phi\right)\, .
\end{equation}
Here, $a=J/M$ where $M$ and $J$ are the ADM mass and angular momentum, while $Q$ and $P$ are the black hole's electric and magnetic charge, defined as
\begin{equation}
Q=\frac{1}{4\pi}\int_{S^{2}}\star F\, , \ \ \ P=\frac{1}{4\pi}\int_{S^{2}} F\, ,
\end{equation}
where $S^{2}$ is any surface of constant $r$. Provided that $M^{2}\geq a^{2}+Q^{2}+P^{2}$, the outer horizon is at\footnote{If $M^{2}<a^{2}+Q^{2}+P^{2}$, then the spacetime exhibits a naked singularity, as discussed in Section \ref{CC}.} 
\begin{equation}
r_{H}= M+\sqrt{M^{2}-(a^{2}+Q^{2}+P^{2})}\, ,
\end{equation}
and it coincides with the Killing horizon of the Killing vector field
\begin{equation}\label{k}
k=\partial_{t}+\Omega_{H}\partial_{\phi},
\end{equation}
where $\Omega_{H}=a/(r_{H}^{2}+a^{2})$ is the angular velocity of the black hole as measured by an observer at infinity. In order to keep electric-magnetic duality manifest, it is useful to collect $(F,\star F)$, $(P,Q)$ and $(g,e)$ in two-component vectors
\begin{equation}\label{FIqI}
F^{I}={\left(\begin{array}{@{}c@{}}F \\ \star F \end{array}\right)}\, ,\ Q^{I}={\left(\begin{array}{@{}c@{}}P \\ Q \end{array}\right)}\, , \ q^{I}={\left(\begin{array}{@{}c@{}}g \\ e \end{array}\right)}\, , \ \ \ \ (I=1,2)\, ,
\end{equation}
and introduce the euclidean and symplectic metrics
\begin{equation}\label{delOm}
   \delta_{IJ}={\left(\begin{array}{@{}cc@{}}1 &0 \\ 0 & 1  \\ \end{array}\right)}\, , \ \ \  \Omega_{IJ}={\left(\begin{array}{@{}cc@{}}0 &1 \\ -1 & 0  \\ \end{array}\right)} \, .
\end{equation}
Electric-magnetic duality transformations are generated by 2-dimensional rotations $S^{I}_{\ J}\in SO(2)$,
\begin{equation}
    S^{I}_{\ J}={\left(\begin{array}{@{}cc@{}}\cos\alpha &\sin\alpha \\ -\sin\alpha & \cos\alpha  \\ \end{array}\right)}\, , \ \ \ \alpha\in\mathbb{R}\, ,
\end{equation}
with respect to which $F^{I},Q^{I},q^{I}$ behave as vectors $V^{I}\to S^{I}_{\ J}V^{J}$. Under these transformations the metrics \eqref{delOm} remain invariant, that is,
\begin{align}\label{dtrafo}
  \delta_{IJ}&=S^{K}_{\ I}S^{L}_{\ J}\delta_{KL}\,, \\
  \Omega_{IJ}&=S^{K}_{\ I}S^{L}_{\ J}\Omega_{KL}\, .
\end{align}
Then, \eqref{GeodEq} takes the form
\begin{equation}\label{GeodEqInv}
u^{a}\nabla_{a}u_{b}=\frac{1}{m}\Omega_{IJ}F^{I}_{ba}q^{J}u^{a}\, 
\end{equation}
which is manifestly duality-invariant. Furthermore, in substituting \eqref{MaxVec} in the right-hand side of \eqref{GeodEqInv} one finds that the equations of motion only depend on the following invariant combinations of the charges\footnote{As well as $P^{2}+Q^{2}$, which is a background quantity.}
\begin{align}
    \delta&=\delta_{IJ}Q^{I}q^{J}=Qe+Pg\, ,\\
    \Omega&=\Omega_{IJ}Q^{I}q^{J}=Pe-Qg\, .
\end{align}
$\delta$ and $\Omega$ can be seen as a measure of the ``electric-electric'' and ``electric-magnetic'' interactions, respectively. In particular, $\Omega=0$ in the case that both the particle and the hole are purely electric, while $\delta=0$ if, say, the hole is purely magnetic and the particle is purely electric (and the same is true for any configuration related to these cases by a duality transformation).

Next, we want to construct quantities that are constant along trajectories satisfying \eqref{GeodEqInv}. This follows if there is a Killing vector field $X$ that leaves invariant the Maxwell field strength, $\pounds_{X}F^{I}=0$. Indeed, to such $X$ one can associate a duality vector
\begin{equation}
\mathcal{P}^{I}_{X}={\left(\begin{array}{@{}c@{}}\mathcal{P}_{X} \\ \tilde{\mathcal{P}}_{X} \end{array}\right)}\, ,
\end{equation}
whose components are functions defined by the equation\footnote{That $\mathcal{P}^{I}_{X}$ exists locally is guaranteed by the fact that $dF^{I}=0$ and $\pounds_{X}F^{I}=0$. If the spacetime is simply-connected then $\mathcal{P}^{I}_{X}$ is also globally defined and unique up to a shift by a constant.}
\begin{equation}\label{mm}
    \nabla_{a}\mathcal{P}_{X}^{I}=-X^{b}F^{I}_{ba}\, ,
\end{equation}
and it then follows that the quantity
\begin{equation}\label{cc1}
C_{X}=u_{a}X^{a}+\frac{1}{m}\Omega_{IJ}\mathcal{P}_{X}^{I}q^{J}
\end{equation}
is constant along trajectories satisfying \eqref{GeodEqInv}. As long as the spacetime is asymptotically flat, it is always possible to chose the asymptotic boundary condition
\begin{equation}\label{boundaryCond}
\int_{S^{2}_\infty}\mathcal{P}^{I}_{X}d\Omega=0\, , 
\end{equation}
where $S_{\infty}^{2}$ denotes an asymptotic 2-sphere, so $\mathcal{P}^{I}_{X}$ is uniquely determined. With this choice, $\mathcal{P}_{X}$ and $\tilde{\mathcal{P}}_{X}$ are the electric and magnetic momentum maps of $X$, following the terminology of \cite{Ortin:2015hya,Elgood:2020svt,Ortin:2022uxa}, which are closely related to the electric and magnetic potentials of the black hole when $X$ is the generator of the event horizon (see Section \ref{CC}). Besides being manifestly duality-invariant, the conserved quantities \eqref{cc1} have the advantage of being independent of the choice of gauge of the Maxwell potential $A_{\mu}$. This property is particularly desirable when considering magnetically charged black holes, where the gauge potentials are necessarily singular even in the exterior of the black hole. Another class of constant of motion might be available if the spacetime exhibits a Stackel–Killing tensor $K_{\mu\nu}$ \cite{GDZPPN002254409,Carter:1973rla}, defined by the properties 
\begin{align}
&K_{ab}=K_{ba}\,, \ \ \nabla_{(a}K_{bc)}=0\,, \\ \label{KS}
&K_{c(a}F_{b)}^{\ c}=0\, , \ \  K_{c(a}\star F_{b)}^{\ c}=0\, .
\end{align}
If such tensor exists, then 
\begin{equation}\label{cc2}
    C=K_{ab}u^{a}u^{b}
\end{equation}
is a constant of motion, as can be readily verified. Applying the discussion above (which is not specific to any solution) to the case of dyonic KN black holes, one finds that there exist four independent constants of motion \cite{Carter:1973rla}. One of them follows from the fact that the metric itself is a Stackel–Killing tensor, and the corresponding constant of motion \eqref{cc2} gives nothing but the usual mass-shell condition for a point particle. There are two other constants of the type \eqref{cc1} and follow from the Killing vector fields $\partial_{t}$ and $\partial_{\phi}$, with associated electric momentum maps
\begin{align}\label{mmKN}
\mathcal{P}_{t}(P,Q)=&-\frac{Q r -P a \cos \theta }{\Sigma }\,\\
\mathcal{P}_{\phi}(P,Q)=&\frac{ Q a r \sin ^2\theta - \left(a^2+r^{2}\right)P \cos \theta}{\Sigma }\,
\end{align}
while the magnetic ones are simply obtained as $\tilde{\mathcal{P}}_{t}(P,Q)=\mathcal{P}_{t}(Q,-P)$ and $\tilde{\mathcal{P}}_{\phi}(P,Q)=\mathcal{P}_{\phi}(Q,-P)$. 
A fourth, less obvious constant follows from the Stackel–Killing tensor \cite{PhysRev.174.1559,Walker:1970un,Carter:1973rla}
\begin{equation}\label{KT}
K_{ab}=2\Sigma l_{(a}n_{b)}+r^{2}g_{ab}\, ,
\end{equation}
where
\begin{align}
l&=\frac{r^{2}+a^{2}}{\Delta}\partial_{t}+\partial_{r}+\frac{a}{\Delta}\partial_{\phi}\, ,\\
n&=\frac{1}{2\Sigma}\left[(r^{2}+a^{2})\partial_{t}-\Delta \partial_{r}+a \partial_{\phi}\right]\, ,
\end{align}
are principal null vectors of a Kinnersly tetrad (we recall that the dyonic KN solution is of Petrov type D, just as the vacuum Kerr solution). The tensor \eqref{KT} has the same form as the usual Killing tensor of Kerr \cite{Walker:1970un} (which may be seen as a consequence of the Killing--Yano tensor of type D metrics \cite{Stephani:2003tm}), and it can be verified that it also satisfies the algebraic condition \eqref{KS}. We choose to work in terms of the energy and angular momentum per unit mass, and a generalisation of the usual Carter constant \cite{PhysRev.174.1559} defined respectively as
\begin{align} \notag
\mathcal{E}&\equiv-u_{a}\left(\partial_{t}\right)^{a}-\frac{1}{m}\Omega_{IJ}\mathcal{P}_{t}^{I}q^{J}\\ \label{E}
&=-u_{a}\left(\partial_{t}\right)^{a}+\frac{r \delta-a \Omega \cos{\theta} }{m\Sigma}\, ,\\ \notag 
\mathcal{L}&\equiv u_{a}\left(\partial_{\phi}\right)^{a}+\frac{1}{m}\Omega_{IJ}\mathcal{P}_{\phi}^{I}q^{J}\\ \label{L}
&=u_{a}\left(\partial_{\phi}\right)^{a}+\frac{ar\delta\sin^{2}{\theta}-\Omega(r^{2}+a^{2})\cos{\theta}}{m \Sigma}\, ,\\ \label{K}
\mathcal{K}&\equiv u^{a}u^{b}K_{ab}-(\mathcal{L}-a\mathcal{E})^{2}\, .
\end{align}
Then, using the Mino--Carter time $d\tau=\Sigma d\lambda$ \cite{Mino:2003yg} as a curve parameter and introducing $z=\cos{\theta}$, the equations of motion can be decoupled and cast in the form 
\begin{align} \notag
\left(\frac{dr}{d\lambda}\right)^2 =& \frac{\left(m(\mathcal{E}(r^2+a^2)-a  \mathcal{L}) - r\delta \right)^2}{m^2}\\ \notag
&-\Delta(r^2+(a\mathcal{E}-\mathcal{L})^2+\mathcal{K})\\ \label{eq:radialeom}
 \equiv & R(r) \, ,
\\ \notag \\ \notag
\left(\frac{dz}{d\lambda}\right)^2 = & \mathcal{K}-z^2 \mathcal{K}- \frac{z^2 (\Omega^2 + m^2 \mathcal{L}^2 )+ 2  m \Omega \mathcal{L} z  }{m^2}\\ \notag
&+ a  z(1-z^2)\frac{2  \mathcal{E} \Omega - a m z (1-\mathcal{E}^2)}{m} \\ \label{eq:polareom}
 \equiv & Z(z) \, ,
\\ \notag \\ 
\label{eq:azimuthaleom}
\frac{d\phi}{d\lambda} =& \frac{\Omega  z + m(\mathcal{L} - a \mathcal{E} (1-z^2))}{m(1 - z^2)}\\ \notag
&- \frac{a}{m \Delta}(\delta r + a m \mathcal{L} - m (a^2 + r^2) \mathcal{E})\, ,
\\ \notag \\ 
\label{eq:timeeom}
\frac{dt}{d\lambda} =& (a^2+r^2)\frac{m (a^2+r^2)\mathcal{E} - a m \mathcal{L} - \delta r}{m \Delta} \\ \notag
& +a\frac{  m (\mathcal{L} - a \mathcal{E} (1-z^2)) +\Omega z}{m}  \, .
\end{align}
Using the same methods outlined and used in \cite{LabanMutrie, Fujita:2009bp, vandeMeent:2019cam, Gralla:2019ceu, Dyson:2023fws} we solve these equations analytically for generic plunging and bound geodesics in terms of Jacobi Elliptic functions, a \texttt{Mathematica} package implementing our solutions has also been made available on \href{https://github.com/ConorDyson/DyonicKNGeodesics}{\texttt{Github}} alongside this paper. Our code is built using the same structure as the KerrGeodesics package of the Black Hole Perturbation Toolkit \cite{BHPToolkit}.  

At this point, it is worth pointing out a few differences with respect to the motion of uncharged particles. First, both the energy and the angular momentum of the particle receive electromagnetic contributions proportional to $\delta$ and $\Omega$. Quite remarkably, if $\Omega\ne0$ the angular momentum is nonzero even if the particle lies along the axis $(\theta=0,\pi)$. This striking feature plays an important role in our work as discussed below, and it is a well known consequence of placing together electric and magnetic charges. Another difference concerns the interpretation of $\mathcal{K}$. In the case of neutral particles, one finds that bound ($\mathcal{E}^{2}<1$) geodesics with $\mathcal{K}=0$ are necessarily confined to the equator $z=0$, which is clearly not true if $\Omega\ne0$, even if the black hole is non-rotating $a=0$. Thus, unless the configuration is purely electric or neutral, $\mathcal{K}$ can no more be interpreted as a measure of off-equatorial motion. However, either charged or neutral particles that hit the curvature singularity (the ring $r^{2}+a^{2}z^{2}=0$) must have $\mathcal{K}=0$ and $\mathcal{L}=a \mathcal{E}$,\footnote{The latter of these conditions is specific of KN, and is not necessary in vacuum Kerr.} as follows immediately by requiring that $R(0)\geq0$ and $Z(0) \geq 0$.

\section{Accretion by Rotating Magnetic Black Holes} \label{AC}

A main goal of this work is studying aspects of accretion of electrically charged matter by magnetic black holes that have no counterpart in the well understood purely electric or neutral cases. Therefore, in this section we focus on black holes with magnetic charge only, $P$, while surrounding particles are assumed to carry only electric charge $e$ (this corresponds to setting $\delta=0$ and $\Omega=Pe$ in the invariant variables introduced above). However, before considering a gravitating system it is worth revisiting one of the more salient features of the electromagnetic field created by an electric charge $e$ and a magnetic charge $g$ put together at rest separated by some distance. Such a system was considered by Thomson back in 1904 \cite{thomson_2009} (and therefore we will refer to it as Thomson's dipole) who pointed out that, in spite of being axially symmetric and static, the electromagnetic field possesses a non-vanishing angular momentum which is independent of the distance between charges. It is given by
\begin{equation}\label{JThomsondipole}
\vec{J}=\frac{1}{4\pi}\int_{R^{3}}\vec{r}\wedge\left(\vec{E}\wedge\vec{B}\right)dx^{3}=-eg\,\,\hat{r}\, ,
\end{equation}
where $\hat{r}$ is the unit vector pointing from the magnetic charge into the electric one. This surprising fact provides a way of obtaining Dirac's quantisation condition by assuming that $\vert \vec{J} \vert$ is quantised in half-integer units of $\hbar$ \cite{PhysRev.75.1968,PhysRev.75.309}, so \footnote{The half-integer, as opposed to just integer quantisation is rather unnatural. However, the angular momentum of a Thomson dipole in a more physical scenario than that of two point charges at rest leads to the right integer quantisation \cite{PhysRev.140.B1407}.}
\begin{equation}
    2ge/\hbar=0,\pm1,\pm2, ... \in Z\, .
\end{equation}
One way of generalising this picture to include gravitation consists in replacing the magnetic point charge $g$ by a magnetic black hole with charge $P$. This was first envisaged in \cite{Garfinkle:1990zx} and independently later on in \cite{Bunster:2007sn}, where the authors considered the process of dropping an electric charge radially into a non-rotating magnetic black hole (see also \cite{Kim:2007ca}). Initially, when the charge and the hole are infinitely far apart the total angular momentum is precisely that of a Thomson dipole \eqref{JThomsondipole}. As the particle falls radially, the electromagnetic field exerts a torque on the hole, which starts spinning. The initial angular momentum keeps being transferred into the hole until, eventually, the particle crosses the horizon and what is left is a dyonic Kerr--Newman black hole with angular momentum equal to that of the initial Thomson dipole. In other words, one can (rather strikingly) spin up a non-rotating magnetic black hole by dropping radially an electric particle into it\footnote{In fact, in \cite{Bunster:2007sn} it was speculated that a fraction of the rotation of nowadays neutral black holes could be due to magnetic charges.}. The main purpose of this section is to understand how the previous picture changes if the magnetic black hole is allowed to possess finite (even maximal) angular momentum, and from that derive what are the most distinctive features of accretion of electric matter by rotating, magnetic black holes.

\begin{figure}[t!]
  \includegraphics[width=8.6cm]{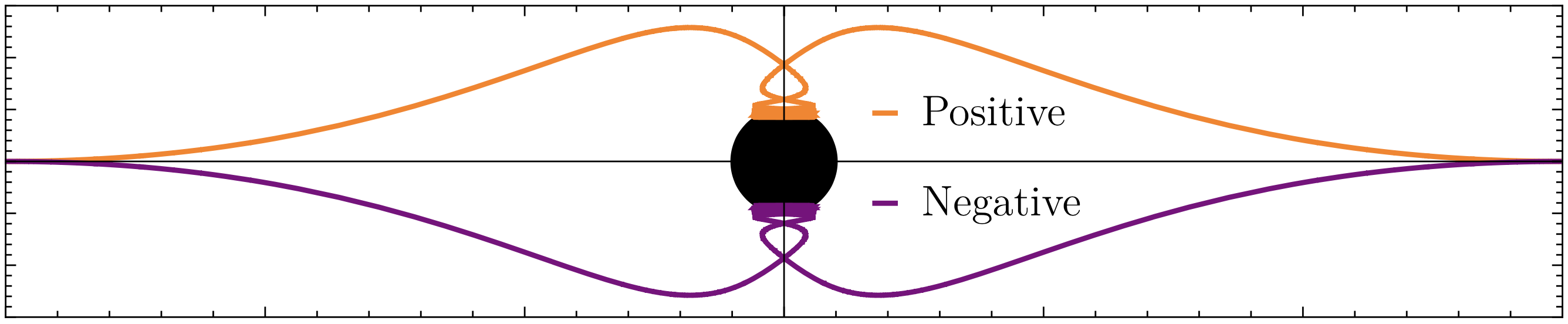}
\caption{\footnotesize{This figure shows the evolution of two pairs of electric particles on a magnetic KN black hole. Each pair has two particles with identical initial conditions at the equator, but opposite charge signs. As discussed in the text, the motion of positive and negative charges are the mirror image of each other with respect to the equator.}}\label{Fig1}
\end{figure}

A first observation is that, even though the metric is invariant under an equatorial $Z_{2}$-transformation, the field strength picks a sign
\begin{equation}\label{Z2Sym}
 F_{\mu\nu}\left(r,\pi-\theta\right)=-F_{\mu\nu}\left(r,\theta\right)\, ,
\end{equation}
which can be seen as a consequence of the dipolar structure of the electric field created by a rotating magnetic charge. This is unlike electric KN black holes, both asymptotically flat and magnetised \cite{Ernst:1976mzr,ErnstWild}, where $F$ is invariant under a $Z_{2}$-transformation. An immediate consequence of this is that, unlike rotating magnetised electric black holes, rotating magnetic ones will not tend to grow a net amount of electric charge when immersed on an ionised homogeneous medium. Indeed, the symmetry \eqref{Z2Sym} implies that the mirror image with respect to the equatorial plane of any solution, $(t(\tau),r(\tau),\theta(\tau),\phi(\tau))\mapsto(t(\tau),r(\tau),\pi-\theta(\tau),\phi(\tau))$, is itself a solution for a particle with opposite charge. Hence, for every plunging trajectory of a particle with charge $e$ there is another one for a particle with charge $-e$ so no net charge is expected to accumulate in the hole (see Figure \ref{Fig1}).
This conclusion also follows by inspection of the black hole's electric potential, $\phi$, which as discussed in Section \ref{CC} is given by
\begin{align}
    \phi&=-\mathcal{P}_{k}\vert_{\mathcal{H}}
\end{align}
where $k=\partial_{t}+\Omega_{H}\partial_{\phi}$ is the Killing vector that generates the event horizon $\mathcal{H}$, and $\mathcal{P}_{k}=\mathcal{P}_{t}+\Omega_{H}\mathcal{P}_{\phi}$ the associated electric momentum map (see \eqref{mmKN}). If the balck hole has no electric charge, as in our case, then $\phi$ vanishes thus indicating that accretion of a net electric charge is not energetically favored. This is in contrast with the case of magnetised electric KN black holes, where imposing the vanishing of the electric potential (the corresponding $\mathcal{P}_{k}$ evaluated at the horizon) requires endowing the hole with certain amount of electric charge \cite{PhysRevD.10.1680}. 

We have established that the black hole tends to accrete the same amount of positive and negative charges, thus remaining globally neutral. Now we shall argue that in such process the black hole reduces its angular momentum. For simplicity, we restrict the discussion to particles whose motion is confined to the axis, $\theta=0,\pi$. Just as a Thomson dipole, the angular momentum per unit mass of such particles is (see \eqref{L})
\begin{equation}\label{angZ}
\mathcal{L}= \biggl\{ {\begin{array}{@{}c@{}}-Pe/m \ \ \ (\text{if} \ \theta=0) \\ +Pe/m \ \ \ (\text{if} \ \theta=\pi) \end{array}} \, ,
\end{equation}
and the Lorentz force per unit mass they are subject to can be written as
\begin{equation}\label{LF}
u^{\mu}\nabla_{\mu}u^{r}=\frac{e}{m}F^{r}_{\ \nu}u^{\nu}=a\mathcal{L}\frac{2r\Delta(r)}{\left(r^{2}+a^{2}\right)^{3}}u^{t}.
\end{equation}
Outside the horizon $\Delta(r)>0$ and a future-directed timelike trajectory has $u^{t}>0$ (see Section \ref{PP}), so the force \eqref{LF} is repulsive or attractive depending on whether $a$ and $\mathcal{L}$ have equal or opposite sign, respectively (although we considered motion along the axis similar reasoning can in fact be applied to a particle at any point in the exterior of the hole).\footnote{It is also natural to wonder whether this spin-spin repulsive force could balance the gravitational attraction between two rotating black holes with comparable masses, one carrying electric charge and the other magnetic one. To the best of our knowledge, non-extremal solutions of this nature need an additional ``dilatonic force'' to prevent the collapse \cite{Chen:2012dr}.} That is, a rotating magnetic black hole tends to accrete charges whose angular momentum differs in sign with that of the hole itself, thus reducing its angular momentum.

In sum, if the magnetic black hole is allowed to rotate the picture one is left with is quite the opposite of that envisaged in \cite{Garfinkle:1990zx,Bunster:2007sn}, where a non-rotating black hole is spun up by absorbing electric charge. Instead, a rotating magnetic black hole immersed on an homogeneous, ionised medium tends to lose its angular momentum by accreting charges of opposite signs (even along the rotation axis, resembling the radial collapse of a Thomson dipole) keeping its net electric charge equal to zero. Reducing the angular momentum of a black hole via accretion is conspicuously reminiscent of the Penrose process, and it is interesting to explore how the latter works if the black hole also possesses magnetic charge. In Section \ref{PP} we consider this issue for magnetic KN black holes and show that there exist novel regions in spacetime from which energy and angular momentum can be extracted. These have no counterpart, either in the mechanical Penrose process \cite{Penrose:1969pc} or in its magnetised version \cite{Wagh:1985vuj}, since they can contain the axis of rotation and can be disconnected from the black hole itself. 

\section{The Penrose Process} \label{PP}

Since its conception \cite{Penrose:1969pc}, the Penrose process has played a prominent role in guiding our intuition when studying dynamical systems involving black holes. In its original and simplest version, it consists of extracting ``rotational'' energy from the hole by mining the so-called ergoregion, a region in the vicinity of the hole where particles are allowed to be in ``negative energy states'' relative to asymptotic observers. The Penrose process generalises naturally to the case that in-falling matter are waves of arbitrary kind, in a phenomenon known as superradiance, whose discovery potential is invaluable (see e.g.~\cite{Brito:2015oca} and references therein). 

At first sight, the Penrose process seems to provide a simple explanation of some high-energy phenomena in which black holes are expected (or known) to be involved, such as active galactic nuclei, relativistic jets and high- and ultra-high-energy cosmic rays. However, it was soon realised \cite{TP,Wald:1974kya} that, in order to be a viable process for extracting energy and angular momentum from the hole, the velocities of the decays (or break ups) at the ergoregion need to be in the relativistic regime $v>1/2$, and in any case the efficiency of the process is bounded to $\lesssim 20 \% $. This is true for the mechanical Penrose process, which only involves a neutral rotating black hole. A much different situation arises if the black hole is immersed on an homogeneous magnetic field \cite{Wald:1974np}, yielding the so-called magnetised Penrose process, which was first envisaged in \cite{Wagh:1985vuj}. In that case, the resulting electric field (due to the twisting of magnetic field lines induced by the hole's rotation) enhances the Penrose process if the particles resulting from a decay are charged. Assuming that the magnetic field is created by reasonable matter orbiting the hole, the break up velocities in the decay need no more be relativistic for the magnetised Penrose process to be viable, and efficiencies can be much larger than the aforementioned $20 \% $ \cite{Tursunov:2019oiq}. However, it is well known that a rotating black hole immersed in an homogeneous magnetic field will accrete a net amount of electric charge, and this turns out to suppress significantly energy extraction \cite{Gupta:2021vww}. Besides, the fact that the magnetised Penrose process relies on having a black hole which is not in isolation makes the system quite difficult to model. \footnote{One simplification consists in regarding the magnetic field as an external one with unknown source, which gives valid predictions in the vicinity of the hole and it is possible to account for the full backreaction by using the Ernst--Will solution \cite{Ernst:1976mzr,ErnstWild}. Another option consists in working perturbatively and model the magnetic field with a specific matter source \cite{Wagh:1985vuj,Dadich2}, which is an asymptotically flat set up but makes it difficult to account for backreaction.} Alternatively, one could enhance energy extraction while keeping the hole in isolation by allowing it to possess a net amount of electric charge. Unfortunately, in that situation the hole would quickly discharge via Schwinger pair creation \cite{Gibbons:1975kk}. 

Here we consider endowing the hole with magnetic charge. This is qualitatively different from the cases discussed above, since energy extraction is greatly enhanced (as shown below) while keeping the black hole in isolation, and no discharge mechanism is expected to neutralise the hole since magnetic monopoles are much less likely to pair create than electric charges. Of course, as discussed in the Introduction the price to pay is the a priori exotic primordial origin of the magnetic charges. In the sake of completeness, we shall derive the main equations for the most general charge configurations first, and then specialise them to the case of a magnetic black hole and electrically charged particles. 

The four-velocity of a particle at a given point can be parametrised using $u^{r},u^{\theta},\mathcal{L}$, while the fourth degree of freedom is fixed using the timelike condition and requiring that the particle's trajectory is future-oriented, which in BL coordinates simply ammounts to imposing $u^{t}>0$.\footnote{This can be seen by defining the so-called zero angular momentum observer $U=-\frac{dt}{\sqrt{-g^{tt}}}$, which is timelike and future-oriented everywhere outside the hole. The requirement that a timelike trajectory $u^{\mu}$ is also future-oriented is $0>U_{\mu}u^{\mu}=-u^{t}/\sqrt{-g^{tt}}$.} Then, the energy per unit mass is no more a free parameter but a function given by

\begin{widetext}
\begin{equation}
\begin{aligned}\label{eq:energysurfac}
\mathcal{E}(u^{r},u^{\theta},\mathcal{L})=&-\frac{1}{m}\Omega_{IJ}\mathcal{P}^{I}_{t}q^{J}-\frac{g_{t\phi}}{g_{\phi\phi}}\left( \mathcal{L}-\frac{1}{m}\Omega_{IJ}\mathcal{P}^{I}_{\phi}q^{J} \right)\\
&+\sqrt{\frac{\Delta \sin^{2}\theta}{g_{\phi\phi}}\left(1+\frac{\left(\mathcal{L}-\frac{1}{m}\Omega_{IJ}\mathcal{P}^{I}_{\phi}q^{J}\right)^{2}}{g_{\phi\phi}}+g_{rr}(u^{r})^{2}+g_{\theta\theta}(u^{\theta})^{2}\right)}\, .
\end{aligned}
\end{equation}
\end{widetext}

It is easy to see that outside the event horizon there are states with $\mathcal{E}<0$ (of course, as in the usual Penrose process, this is not in contradiction with having positive kinetic energy with respect to a local inertial observer). We want to find the regions of spacetime where particles with a given angular momentum $\mathcal{L}$ \textit{can} be in a negative energy state, since it is in those regions where decays or break ups could lead to energy and angular momentum extraction. From \eqref{eq:energysurfac}, it is clear that the minimal energy states are those with $u^{r}=u^{\theta}=0$. So, with that choice, the zero-energy level sets given by \eqref{eq:energysurfac} enclose the regions where negative energy states are allowed. 

\begin{figure*}[t!]
  \includegraphics[width=18cm]{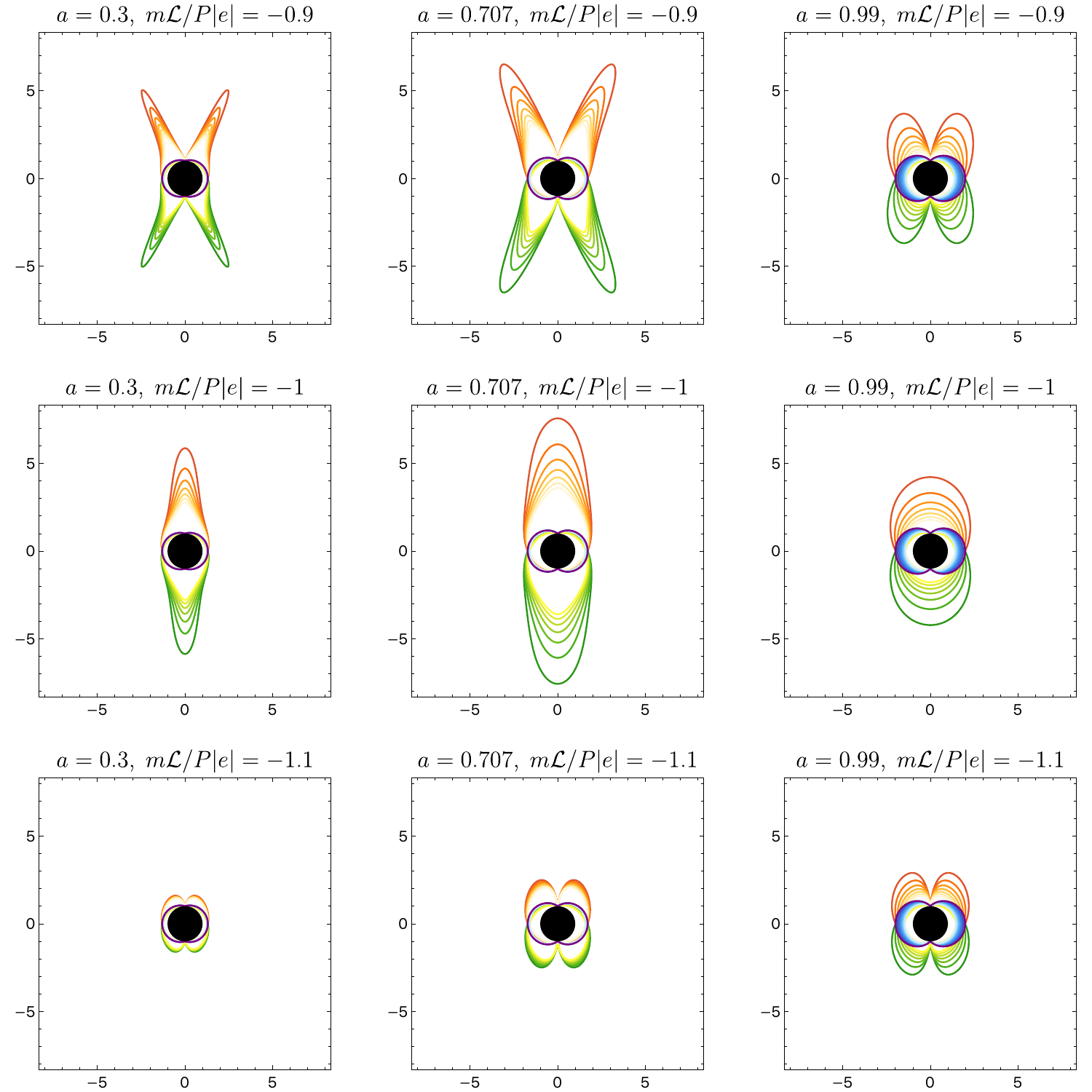}
\caption{Regions of negative energy states of an electric particle with $e/m=\pm 100$ (red for positive, green for negative) in a rotating magnetic black hole with $M=1$, $Q=0$, $\epsilon=10^{-3}$. The columns correspond to the spin parameters $a=0.3,0.707,0.99$ and the corresponding positive value of magnetic charge $P$ determined by \eqref{extr}, while rows display different values of the particle's angular momentum $m\mathcal{L}/P \lvert e \rvert =-0.9,-1,-1.1$. The contours show the regions of negative energy states given by \eqref{eq:energysurfac} with $u^{r}=u^{\theta}=0$ (see text). The outermost contour is the zero energy level, and subsequent inner curves decrease by $\Delta \mathcal{E}  = -0.5$ the value of the energy level, following the colour scale. In the same conventions, blue contours correspond to a neutral particle, so they are associated to the mechanical Penrose process.}\label{3by3}
\end{figure*}

Let us focus on the case of an electric particle with charge $e$ and a purely magnetic black hole of charge $P$, so $Q=g=0$. We fix the overall scale by setting $M=1$ and introduce the extremality parameter  
\begin{equation}\label{extr}
    \epsilon=\sqrt{1-a^{2}-P^{2}}\, ,
\end{equation}
so $\epsilon=0$ for extremal black holes and $\epsilon=1$ for neutral, non-rotating ones. We find negative energy states for values of $\mathcal{L}$ with opposite sign to that of $a$, similarly to the mechanical Penrose process. There are two regimes of $\mathcal{L}$, defined by the value of the angular momentum of the Thomson dipole (see Figure \ref{3by3}):
\begin{itemize}
    \item $0<\vert \mathcal{L}\vert<\vert Pe/m\vert$ : in this regime, the presence of the magnetic charge deforms the region of negative energy states by enlarging it along the directions determined by certain conjugate values of the axial angle, $\theta_{0}$ and $\pi-\theta_{0}$, with $\theta_{0}\in(0,\pi/2)$. $\theta_{0}$ goes from the equator $\theta_{0}=\pi/2$ for $\vert \mathcal{L}\vert \approx 0$ (where there are no negative energy states) and approaches the rotation axis $\theta_{0}=0$ as $\vert \mathcal{L}\vert\to \vert Pe/m\vert$.

    \item $\vert \mathcal{L}\vert=\vert Pe / m \vert$ : when the angular momentum is precisely that of the Thomson dipole, we find that the region of negative energy states includes the rotation axis. This leads to the quite remarkable possibility of extracting angular momentum (and of course energy) from a hole in a process that is entirely axisymmetric (e.g.~a decay happening along the axis). This condition is also the requirement one finds for the existence of motion on the axis.

    \item $\vert \mathcal{L}\vert>\vert Pe/m\vert$ : in this case we find a similar situation to that of the first regime, where now $\theta_{0}\to\pi/2$ as $\vert\mathcal{L}\vert\to\infty$, and the region of negative energy states converges to the mechanical ergoregion. This is as expected, since for large $\mathcal{L}$ at fixed $a$, $P$ and $e$ the mechanical effects dominate over the electromagnetic ones.
\end{itemize}
For all values of $\mathcal{L}$, we notice that the region of negative energy states at the equator $\theta=\pi/2$ is unaffected by $P$, since the momentum maps vanish there (we recall that in this discussion the electric black hole charge is set to $Q=0$)
\begin{equation}\label{noMM}
\mathcal{P}_{t}(\theta=\pi/2)=\mathcal{P}_{\phi}(\theta=\pi/2)=0\, .
\end{equation}
On the other hand, for $\vert \mathcal{L} \vert = \vert P e / m \vert$ there exist negative energy states along the rotation axis. Such states do not exist in the mechanical case, and furthermore carry angular momentum. Expanding in powers of $e/m$ (which may be motivated by the fact that for an electron $e/m\approx2 \times 10^{21}$) one finds that the negative energy states along the axis extend up to
\begin{equation}
   r_{\text{max}} \approx  \sqrt{\lvert aPe/m \rvert } \, ,
\end{equation}
which is largest when $aP$ is maximised (as expected since the magnitude of the electric field is roughly given by $aP$). For a fixed extremality parameter $\epsilon$, this happens at $a=P=\sqrt{(1-\epsilon^{2})/2}$, so for small $\epsilon$ one has $a=P\approx 0.707$. The magnification of the region of negative energy states for that choice of parameters can be clearly seen in Figure \ref{3by3}.  

\begin{figure}[t!]
\centering
 \includegraphics[width=8cm]{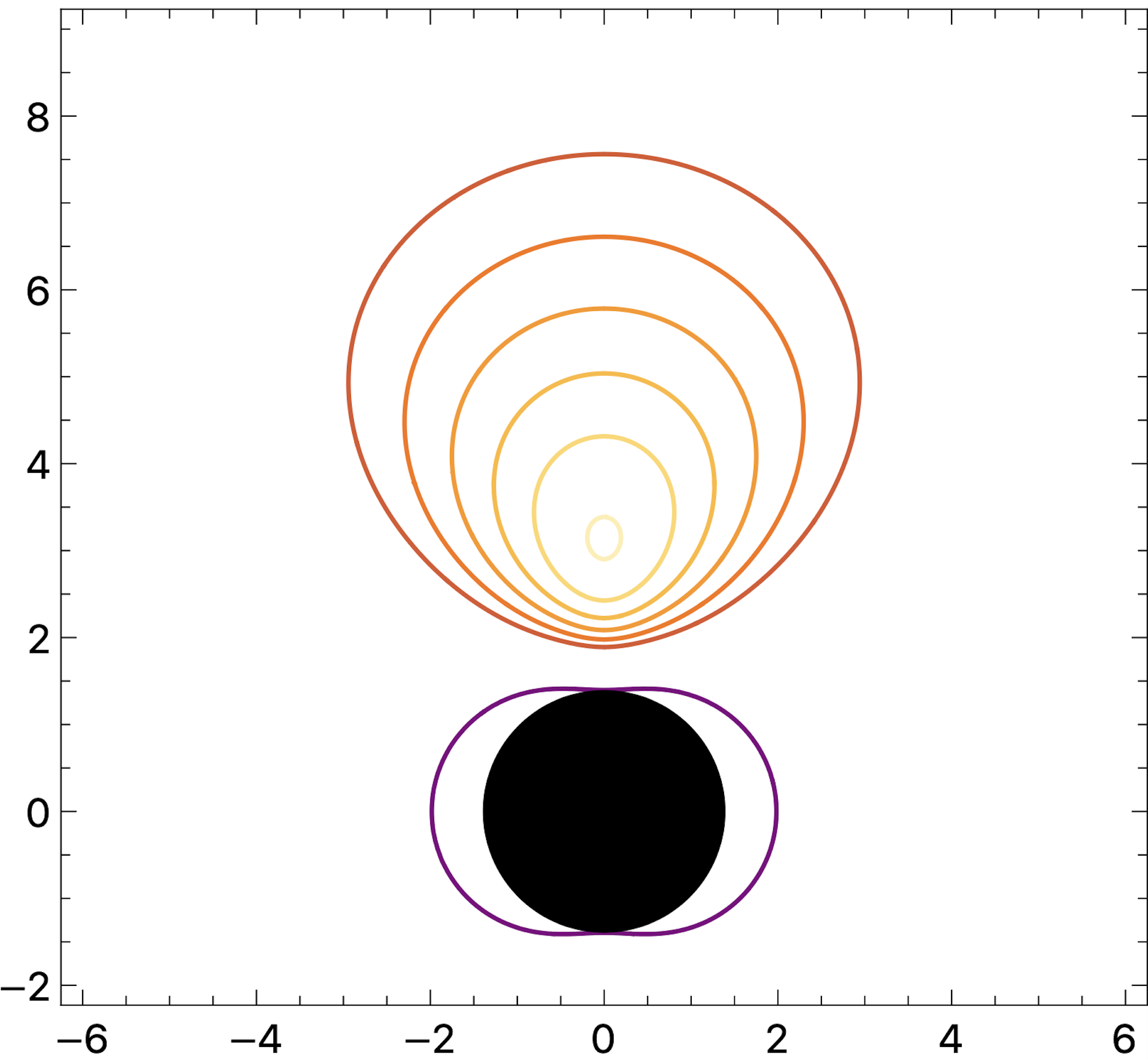}
\caption{Region of negative energy states for a black hole with parameters $M=1$, $a=0.9$, $Q=-0.085$ and $P=-0.16$. The particle's charge to mass ratio is $e/m=100$ and its angular momentum per unit mass is $\mathcal{L}=-Pe/m=16$. The outermost red contour corresponds to the zero-energy surface and the successive inner ones decrease the energy by $-0.1$, following the color scale. The purple contour is the mechanical ergosurface. As explained in the text, the attractive gravitational and coulomb forces are compensated by the repulsive dipolar force, thus leading to disconnected regions of negative energy states.}\label{trappedregions}
\end{figure}

We shall also comment on the case that the black hole is endowed not only with magnetic but also electric charge\footnote{ In the absence of an external magnetic field the hole is expected to quickly lose its electric charge, although some amount could be retained via the Witten effect \cite{Witten:1979ey}.} (then the particle can be chosen to be purely electric without loss of generality). The monopolar electric field enhances the Penrose process and allows both energy and charge extraction even if the black hole is nonrotating. If, in addition, the black hole rotates and possesses magnetic charge the electric field picks a dipolar piece which, in the neighbourhood of one of the components of the rotation axis ($\theta=0$ or $\theta=\pi$) opposes to the monopolar one. Interestingly, this balance of electric fields leads to the existence of ``floating'' regions of negative energy states (see Figure \ref{trappedregions}). That is, after a decay one of the products can reach a negative energy state which is an orbit confined to a neighbourhood of the axis and that never crosses the horizon. Even though that particle never falls into the hole, the other product of the decay can reach infinity and energy is extracted from the system (in such process there is also extraction of electric charge, but not of angular momentum). Regions of negative energy states that are disconnected from the horizon were also found recently in \cite{Gupta:2021vww} in the case of rotating magnetised black holes (i.e.~rotating black holes immersed in an external magnetic field). Those are toroidal regions centered around the hole and symmetric with respect to the equator, while the ones found here associated to magnetic black holes are simply connected bubble-shaped regions and centered at a point of the axis. 
Particles trapped in negative energy regions which are disconnected from the horizon are expected to release more energy via synchrotron radiation, and follow an evolution driven by electromagnetic radiation-reaction \cite{Santos:2023uka,Baker:2023gdc}. We leave a detailed study of this issue for future work.

We conclude this section with some remarks about the bounds on the velocity of the break up and the efficiency of the Penrose process. For completeness, we shall do so for the most general charge configuration of both the hole and the particle. A simple computation shows that the specific energy $\mathcal{E}$ of a particle with mass $m$ and charge $q^{I}$ that is the product of a decay of a particle with specific energy $\mathcal{E}_{0}$, mass $m_{0}$ and charge $q_{0}^{I}$ must satisfy \cite{Wald:1974kya,Wagh:1985vuj}
\begin{widetext}
   \begin{align}\notag
&-\frac{1}{m}\Omega_{IJ}\mathcal{P}^{I}_{t}q^{J}+\gamma(v)\left[\mathcal{E}_{0}+\frac{1}{m_{0}}\Omega_{IJ}\mathcal{P}^{I}_{t}q^{J}_{0}-v\sqrt{g_{tt}+\left(\mathcal{E}_{0}+\frac{1}{m_{0}}\Omega_{IJ}\mathcal{P}^{I}_{t}q^{J}_{0}\right)^{2}}\right]\\ \notag
&\leq \mathcal{E}\\ \label{generalBound}
&\leq-\frac{1}{m}\Omega_{IJ}\mathcal{P}^{I}_{t}q^{J}+\gamma(v)\left[\mathcal{E}_{0}+\frac{1}{m_{0}}\Omega_{IJ}\mathcal{P}^{I}_{t}q^{J}_{0}+v\sqrt{g_{tt}+\left(\mathcal{E}_{0}+\frac{1}{m_{0}}\Omega_{IJ}\mathcal{P}^{I}_{t}q^{J}_{0}\right)^{2}}\right]\, ,
\end{align} 
\end{widetext}
where $v$ is the absolute value of the velocity of the product in the frame of the decaying particle and $\gamma(v)=1/\sqrt{1-v^{2}}$ the Lorentz factor. Applying \eqref{generalBound} to the decay of a neutral particle into electric charges along the rotation axis of a magnetic black hole, one finds that the lower bound can be negative (and therefore energy extraction is actually possible) only if $v$ satisfies
\begin{equation}
v>\frac{1-\alpha^{2}}{1+\alpha^{2}}\, , \ \ \ \ \ \ \ \alpha\equiv\frac{(e/m)aP}{(e_{0}/m_{0})aP+\left(r_{H}^{2}+a^{2}\right)\mathcal{E}_{0}}\, .
\end{equation}
In particular, $v$ can be arbitrarily close (or equal) to zero if $\lvert \alpha \rvert \geq1$. This is true for a process taking place along the rotation axis, and similar conclusions can be deduced for processes in the enlarged regions of negative energy states. As noted above, one exception are processes happening at the equator $\theta=\pi/2$, where \eqref{noMM} holds and the bounds on $v$ are the same as in the mechanical Penrose process. Similarly, it is easy to see from \eqref{eq:energysurfac} and \eqref{generalBound} that the efficiency of energy extraction $\eta\equiv(m\mathcal{E}-m_{0}\mathcal{E}_{0})/m_{0}\mathcal{E}_{0}$ can be made significantly larger than that of the mechanical Penrose process for decays happening off the equator, while at the equator the bounds on the efficiency remain the same. Nevertheless, an astrophysically meaningful discussion about the bounds on both $v$ and $\eta$ requires having some expected values for $P/M$ and $P/J$, an issue that lies beyond the scope of our work and that is left for future research.         

Finally, we remark that similarly to the mechanical Penrose process the amount of energy that can be extracted from the hole is bounded by the irreducible mass, defined as $M^{2}_{IRR}=A_{H}/16\pi=(a^{2}+r_{H}^{2})/4$ where $A_{H}$ is the area of a spatial section of the horizon. Indeed, in a Penrose process involving the decay of a charged particle one has $\delta M_{IRR}\geq0$ (in agreement with the second law of black hole mechanics). At the same time, 
\begin{equation}
    M^2 = \left(M_{IRR}+\frac{P^{2}+Q^2}{4 M_{IRR}}\right)^2 + \frac{J^2}{4 M_{IRR}^2}\, \geq M_{IRR}^{2},
\end{equation}
so the amount of energy that can be extracted from the hole is necessarily smaller than $M-M_{IRR}$.

\section{Cosmic Censorship} \label{CC}

In this last section, we approach a more fundamental question, that of whether accretion of electric matter by magnetic black holes could lead to the formation of a naked singularity, thus incurring into a contradiction with the weak cosmic censorship \cite{Penrose:1969pc}. In general, understanding whether the latter is true given some reasonable physical assumptions stands as one of the major challenges in gravitational physics. While it has been proven rigorously for specific choices of matter fields and under certain symmetry assumptions \cite{C}, a proof of sufficient generality is still elusive. A less formal, yet physically illuminating approach consists in testing the conjecture via a gedankenexperiment in which one tries to overcharge or overspin an extremal BH (say, an extremal electric KN black hole) via a physical process, such as accretion of a (probably fine-tuned) matter wave. Wald considered this in \cite{Wald1974GedankenET} for electric KN black holes in the case that the in-falling matter is point-like but is allowed to possess electric charge as well as internal spin. The work concludes robustly that no naked singularity can form by such accretion process, a fact that can be interpreted intuitively as due to the electric force and the centrifugal barrier preventing the absorption of matter that would overcharge or overspin the hole. Subsequently, the authors in \cite{Natario:2016bay} were able to construct an elegant argument to extend those conclusions to the case of arbitrary in-falling matter, as long as the null-energy condition is respected. This argument was later formalised and extended to higher orders in the perturbations in \cite{Sorce:2017dst}, which allowed the authors to prove that cosmic censorship is also not violated in gedankenexperiment where the hole is arbitrarily close to (but not quite at) extremality. \footnote{Quite crucially, this implies that in gedankenexperiments of the type proposed in \cite{Hubeny:1998ga} it is necessary to perform a complete, second order computation (see \cite{Colleoni:2015ena,Sorce:2017dst} and references therein for a brief overview on this issue).}

However, the works \cite{Wald1974GedankenET,Natario:2016bay,Sorce:2017dst} mentioned above consider only purely electric scenarios (i.e.~electrically-charged black holes accreting electric matter, and configurations related to this by electric-magnetic duality) so their conclusions do not apply, in principle, to the case we are concerned with in this work, which is that of a magnetic black hole accreting electrically charged matter. In fact, Lorentz forces in such set up can be very different from those in a purely electric case, and they might even be exactly vanishing as in the case of an electric charge falling radially into a magnetic non-rotating black hole (the magnetic Reissner--Nordström (RN) black hole). In addition, we have also shown that there are novel ways of inducing rotation into the hole which are drastically different from just transference of orbital angular momentum. These observations make worth exploring whether rotating magnetic black holes are safe from developing a naked singularity. In this discussion, we shall restore full generality and consider extremal dyonic black holes, as well as dyonic in-falling matter.

Given an extremal dyonic black hole, the question is whether an in-falling fluctuation can be such that the induced change on the black hole parameters oversaturate the extremality bound, that is, whether the perturbation can violate the inequality
\begin{equation}\label{boundvio}
(M^{2}+a^{2})\delta M\geq M\left(Q\delta Q+P\delta P\right)+a\delta J\, .
\end{equation}

Hiscock \cite{HGed} was the first (to the best of our knowledge) to address this issue in the case of point-like in-falling charges.\footnote{This problem was reconsidered independently later on in \cite{Semiz:1990fm}, reaching the same conclusions.} Following the general strategy devised in \cite{Wald1974GedankenET}, it was shown that a violation of the weak cosmic censorship cannot occur. Although the proof is satisfactory and general (within the class of matter considered), it is illustrative to discuss in more detail a particular process in order to gain intuition about what mechanism protects cosmic censorship in this case. Let us restrict to an extremal rotating magnetic black hole (so $Q=0$), and consider dropping through the axis ($\theta=0$) an electric charge that is initially at rest at infinity, which fixes $\mathcal{E}=1$. As discussed above (see \eqref{angZ}), such particle carries an angular momentum given by $\mathcal{L}=-Pe/m$. Choosing the sign of the charge so that $\mathcal{L}a>0$, the particle will spin up the hole, if accreted, and potentially violate the extremality bound. Notice that this way of inducing angular momentum into the hole is qualitatively different from the usual transference of orbital angular momentum, since it only involves motion along the axis so no centrifugal barrier prevents the particle from approaching the horizon. However, in \eqref{LF} we showed that the hole's electric field produces a spin-spin repulsion, so a particle that would spin up the hole feels a repulsive force. Particles that could violate cosmic censorship in this process by entering the hole have
\begin{equation}
    1 < \frac{a}{M^{2}+a^{2}}\mathcal{L}\, ,
\end{equation}
but using \eqref{eq:radialeom} it is easy to show that in that case the trajectory has a turning point outside the horizon.\footnote{The turning point would be precisely at the horizon if $ 1 = \frac{a}{M^{2}+a^{2}}\mathcal{L}$, which preserves the extremality condition.} Explicitly said, the spin-spin repulsion prevents from crossing the horizon particles that would contradict \eqref{boundvio} by overspinning the hole.

The case of in-falling matter consisting of a charged scalar field was considered in \cite{Semiz:2005gs} and later on in \cite{Toth:2012vvy}, and neutral and charged Dirac fields were considered in \cite{Duztas:2014sga} and \cite{Toth:2015cda}, respectively. While in the scalar case one reaches the same conclusions as for particle matter, it turns out that Dirac fields could make the black hole turn into a naked singularity. As argued in \cite{Toth:2015cda}, however, this is due to the fact that the Dirac field does not satisfy the null-energy condition (an artifact of regarding Dirac's equation as describing a classical field). 

Our purpose in this section is to extend the above results to the most general set up within the Einstein--Maxwell theory, that is, that of an extremal dyonic KN black hole accreting matter of arbitrary kind. We shall do so by following the strategy envisaged in \cite{Natario:2016bay} and \cite{Sorce:2017dst}, and implementing the covariant phase space techniques introduced in \cite{Ortin:2022uxa} (building on previous works \cite{Elgood:2020svt,Elgood:2020mdx,Elgood:2020nls,Ortin:2021ade,Mitsios:2021zrn}) which allows one to account for both electric and magnetic type contributions to the conserved charges in a gauge- and duality-invariant guise (see also \cite{Meessen:2022hcg,Ballesteros:2023iqb,Gomez-Fayren:2023wxk,Bandos:2023zbs} for generalisations that also account for more general notions of charges, including scalar and Fermionic ones). 

Consider a fluctuation that arises as a solution of the Einstein--Maxwell equations (linearised around an exact electrovacuum solution, such as the dyonic KN black hole) sourced by a linear dyonic current,
\begin{equation}\label{pertnonfixedT}
\begin{aligned}
\delta(G_{ab}-T^{EM}_{ab})&=8\pi T_{ab}\\ 
d\delta F^{I}&=-4\pi \star j^{I}
\end{aligned}
\end{equation}
where $\delta F^{I}$ is the linear variation of the field strength vector \eqref{FIqI}, 
\begin{equation}\label{jI}
j^{I}={\left(\begin{array}{@{}c@{}}j_{(g)} \\ j_{(e)} \end{array}\right)}
\end{equation}
is the dyonic density current. Now, assume that the background spacetime contains a stationary and axisymmetric black hole, whose event horizon is a Killing horizon of a Killing vector field $k=\partial_{t}+\Omega_{H}\partial_{\phi}$ for some constant $\Omega_{H}$, and consider a three-dimensional surface $\Sigma$ that extends from a spacelike 2-sphere at the horizon $S_{\mathcal{H}}^{2}$ to an assymptotic 2-sphere $S_{\infty}^{2}$. By exploiting the symmetries of the theory at hand, one can construct a fundamental identity that any such fluctuation must satisfy on $\Sigma$ \cite{Wald:1993nt,Iyer:1994ys}. In \cite{Ortin:2022uxa} it was understood how to include magnetic type contributions in a gauge- and duality-invariant guise, and here we have extended the identity by including source terms (a detailed derivation in the metric formulation of gravity, as opposed to the vielbeine one used in \cite{Ortin:2022uxa}, can be found in Appendix \ref{VI}). It reads
\begin{widetext}
    \begin{equation}\label{TINTMAST}
\delta M-\Omega_{H}\delta J=\phi^{I}\Omega_{IJ}\delta Q^{J}-\int_{S^{2}_{\mathcal{H}}}\left[\delta \bold{Q} _{k}^{GR}+\iota_{k}\boldsymbol{\Theta}^{GR}\right]+\int_{\Sigma}\left[\mathcal{P}_{k}^{I}\Omega_{IJ}\star j^{J}-k_{a}T^{ab}\boldsymbol{\epsilon}_{b}\right].
\end{equation}
\end{widetext}
Here, $\delta M$ and $\delta J$ are the variations of the ADM mass and angular momentum induced by the fluctuation, 
\begin{equation}
\phi^{I}=-\mathcal{P}_{k}^{I}\lvert_{\mathcal{H}}\,\, , \ \ \ \ \delta Q^{I}=\frac{1}{4\pi}\int_{S_{\mathcal{H}}^{2}}\delta F^{I}
\end{equation}
are the electromagnetic potentials and the variation of the charges of the hole enclosed by $S^{2}_{\mathcal{H}}$, respectively, and $\bold{Q} _{k}^{GR}$ and $\boldsymbol{\Theta}^{GR}$ are the GR's Noether--Wald charge and symplectic potential (whose form is given in Appendix \ref{VI} but is not needed here). As an example of application, for a vacuum fluctuation the last integral on the right hand side of \eqref{TINTMAST} vanishes while the first one gives $(\kappa/8\pi)\delta A_{H}$ \cite{Iyer:1994ys}, so one finds the first law of black hole mechanics where the first term in the right hand side are gauge-invariant work terms of both electric and magnetic types in a duality-invariant combination.   

\begin{figure}[t!]
    \centering
    \includegraphics[scale=0.25]{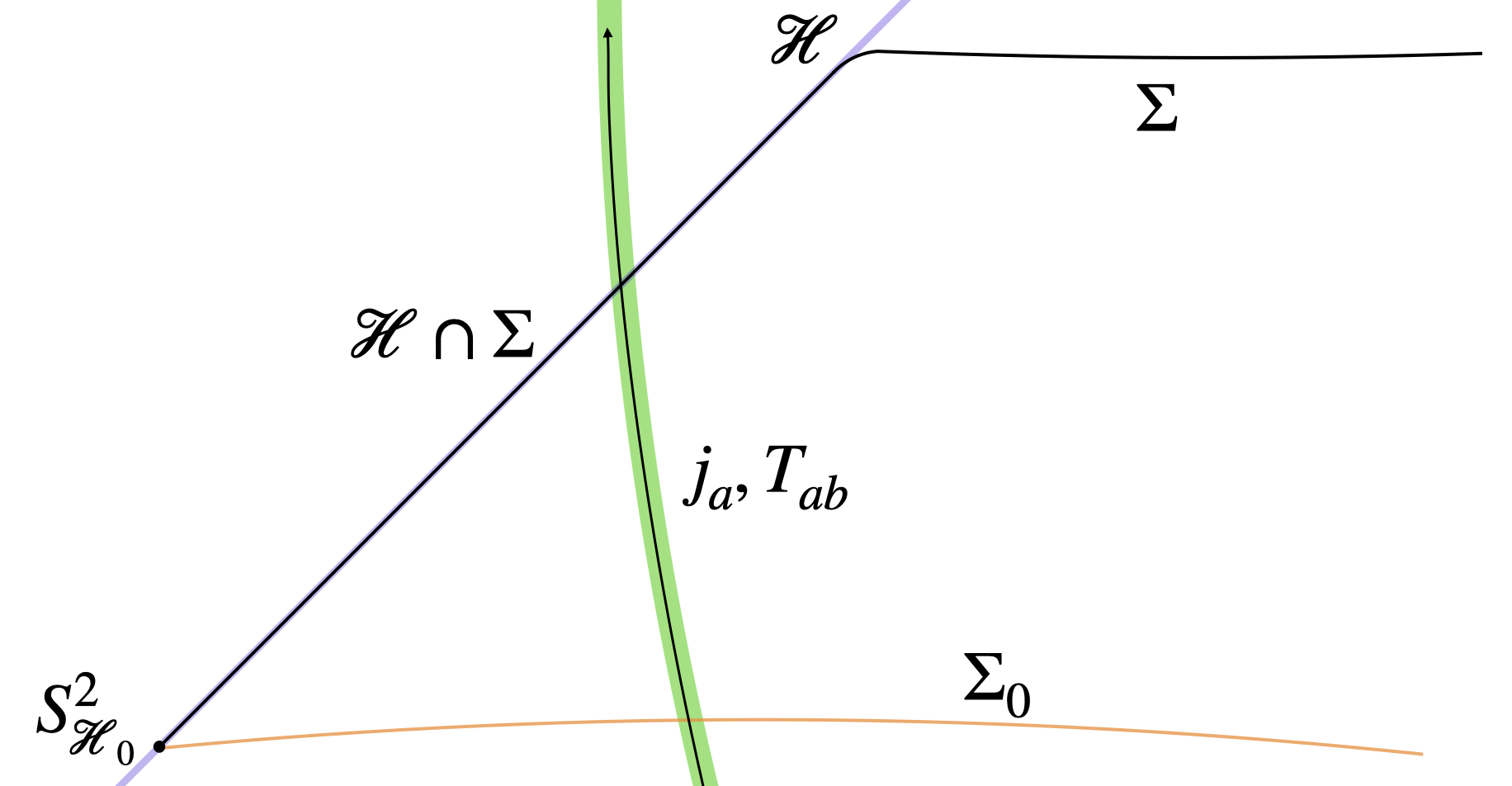}
        \caption{\footnotesize{The horizon $\mathcal{H}$ is in purple, the initial Cauchy slice $\Sigma_{0}$ in orange, $\Sigma$ is in black, with its inner boundary denoted by a black dot, and the source's trajectory is in green.}}
    \label{fig}
\end{figure}

At this point we can follow \cite{Sorce:2017dst} and define our fluctuation from its initial data on a Cauchy slice $\Sigma_{0}$ that extends from a 2-sphere at the horizon $S^{2}_{\mathcal{H}_{0}}$ to infinity. We take sources that are compactly supported and initially far enough from the black hole, so that the fluctuation vanishes at $S^{2}_{\mathcal{H}_{0}}$, and assume for simplicity that all of the matter eventually falls inside the black hole. The 3-surface $\Sigma$ on which we will apply the fundamental identity \eqref{TINTMAST} is taken as follows: it starts at $S^{2}_{\mathcal{H}_{0}}$ and extends along the horizon until all of the matter has crossed it. Then it becomes spatial and extends all the way to infinity (see Fig.~\ref{fig}). 

Since the fluctuation vanishes at $S^{2}_{\mathcal{H}_{0}}$, the first term and the first integral in the right hand side of \eqref{TINTMAST} vanish and one is left with 
\begin{align}\notag
\delta M-\Omega_{H}\delta J&=\int_{\Sigma\cap \mathcal{H}}\left[\mathcal{P}_{k}^{I}\Omega_{IJ}\star j^{J}-k_{a}T^{ab}\boldsymbol{\epsilon}_{b}\right]\\
&=-\phi^{I}\Omega_{IJ}\int_{\Sigma\cap \mathcal{H}}\star j^{J}-\int_{\Sigma\cap \mathcal{H}}k_{a}T^{ab}\boldsymbol{\epsilon}_{b}\\
&=\phi^{I}\Omega_{IJ}\delta Q^{J}_{flux}-\int_{\Sigma\cap \mathcal{H}}k_{a}T^{ab}\boldsymbol{\epsilon}_{b}
\end{align}
where in the second line we used the zeroth law of the electromagnetic field (i.e.~that $\phi^{I}=-\mathcal{P}^{I}_{k}\lvert_{\mathcal{H}}$ is constant on $\mathcal{H}$, see \cite{Ortin:2022uxa} and Appendix \ref{VI}), and in the third line we used that $\delta Q^{I}_{flux}=-\int_{\Sigma\cap \mathcal{H}}\star j^{I}$ is the net amount of charge that has crossed the horizon (see \eqref{pertnonfixedT}). Finally, 
\begin{align}\notag
-\int_{\Sigma\cap \mathcal{H}}k_{a}T^{ab}\boldsymbol{\epsilon}_{b}=\int_{\Sigma\cap \mathcal{H}}V^{a}k^{b}T_{ab}\tilde{\boldsymbol{\epsilon}}
\end{align}
where $V^{a}$ is future-directed and normal to the horizon, so it is proportional to $k^{a}$, and $\tilde{\boldsymbol{\epsilon}}$ is the volume form on $\mathcal{H}$ \cite{Gao:2001ut}. That is, the right hand side is manifestly non-negative if $T_{ab}$ satisfies the null-energy condition. If the latter holds, one arrives at
\begin{align}\notag
\delta M-\Omega_{H}\delta J-\phi^{I}\Omega_{IJ}\delta Q^{J}_{flux}\geq0\, .
\end{align}
Identifying $\delta Q^{I}_{flux}$ with the total amount of accreted charge $\delta Q^{I}$, and evaluating the formula on an extremal dyonic KN background one arrives precisely at \eqref{boundvio}, thus showing that no violation of the weak cosmic censorship takes place if the energy momentum tensor of the in-falling matter satisfies the null-energy condition.

\section{Discussion} \label{D}

In this work we have considered the accretion of charged test matter by rotating dyonic black holes. We have uncovered a rich phenomenology, where the interaction of the angular momentum contained in the electromagnetic field and the spin of the hole plays a fundamental role. Focusing on magnetic black holes, we have shown that when immersed on an ionised medium the hole tends to lose its angular momentum by accreting charges, but that it does so while remaining globally neutral. We have also shown that accretion can happen in a superradiant manner, and that energy and angular momentum can be extracted from the hole via a Penrose process that is greatly enhanced due to the dipolar electric field created by the rotating magnetic charge of the hole. The regions that can accommodate negative energy states extend much further than the mechanical ergoregion, and may even contain the rotation axis or be disconnected from the hole. Finally, we have addressed the issue of whether extremal dyonic rotating black holes can be overcharged or overspun via matter accretion, and have provided a proof that answers it in the negative as long as the in-falling matter respects the null-energy condition, but is otherwise completely general.

It would be interesting to analyse in greater detail the observational potential of some of the remarkable phenomena that magnetic black holes lead to. Besides the Penrose process in its original version discussed here, which involves the decay or disruption of in-falling matter, energy extraction via the products of a collision (the ``collisional'' Penrose process \cite{CPP}) can also lead to exciting phenomenology \cite{Banados:2009pr,Nemoto:2012cq, Harada:2014vka} that is worth understanding in the case that the black hole is endowed with some magnetic charge. It would also be interesting to study superradiant phenomena beyond particle matter. To that end, our equations in Section \ref{CC} allow a fully gauge- and duality-invariant analysis for arbitrary in-falling matter without the need of reducing the equations of motion to decoupled master ODE's (which is not always possible anyways), along the lines of \cite{Toth:2015cda}. Finally, working out the gravitoelectromagnetic waveforms resulting from coalescences involving electric and magnetic black holes is of great interest for multi-messenger astronomy. The Newtonian regime allows an interesting analytic approach to this problem \cite{Liu:2020bag,Liu:2020cds,Liu:2020vsy,Liu:2022wtq,Juraeva:2021gwb}, but future space-based detectors such as LISA also motivate considering extreme mass ratios \cite{Pereniguez:2023wxf}, where both relativistic and strong field effects are crucial. Work along these lines is on the way.

\vspace{1cm}
\noindent {\bf Acknowledgments.} 
 
We would like to thank Vitor Cardoso, Gregorio Carullo, Marc Casals, João Costa, Miguel Montero, Jaime Redondo-Yuste, Tomás Ortín, Maarten Van De Meent and Rodrigo Vicente for interesting and fruitful conversations. We acknowledge financial support by the VILLUM Foundation (grant no. VIL37766) and the DNRF Chair program (grant no. DNRF162) by the Danish National Research Foundation.

\clearpage

\appendix

\onecolumngrid

\section{Fundamental Variational Identity of Dyonic Black Holes}\label{VI}

 In this appendix we provide a detailed derivation of the identity \eqref{TINTMAST} (we follow the notation introduced in \cite{Wald:1993nt,Iyer:1994ys}). Consider the Einstein--Maxwell Lagrangian,
\begin{equation}\label{lagrangian}
\bold{L}[g,A]=\frac{1}{16\pi}\left(R-F^{2}\right)\boldsymbol{\epsilon},
\end{equation}
which is both covariant and gauge-invariant. We proceed by regarding $A_{a}$ as a local 1-form on spacetime, instead of a global connection on a $U(1)$-bundle. However, our equations do not make any assumption on the gauge of $A_{a}$ and, as shown below, this approach allows one to include magnetic-type terms which have been missed, or are not defined, in more general treatments of theories on principal fiber bundles \cite{Prabhu:2015vua}. The linear automorphisms of this theory (essentially, spacetime diffeomorphisms and internal gauge transformations) can be parametrised by a vector field $\xi^{a}$ and a function $\mathcal{P}$, and their action on $g_{ab}$ and $A_{a}$ reads
\begin{equation}\label{gauge}
\delta_{\xi}g_{ab}=-\pounds_{\xi}g_{ab}, \ \ \ \ \ \delta_{\xi,\mathcal{P}}A=-(\iota_{\xi}F+d\mathcal{P}).
\end{equation}
Equivalently, in terms of the gauge parameter
\begin{equation}
\chi(\xi,A,\mathcal{P})\equiv\iota_{\xi}A-\mathcal{P}
\end{equation}
the action on $A_{a}$ takes the more familiar form  $\delta_{\xi,P}A=-\pounds_{\xi}A+d\chi$. Eventually we will be setting $\xi^{a}=k^{a}$ and $\mathcal{P}=\mathcal{P}_{k}$, where $k^{a}$ and $\mathcal{P}_{k}$ are the background quantities corresponding to the Killing generator of the horizon and the electric momentum map, respectively. However, we proceed by first deriving all the identities associated to the symmetry \eqref{gauge} off-shell, for arbitrary $\xi^{a}$ and $\mathcal{P}$. Writing the general first variation of the Lagrangian as
\begin{equation}\label{1stvar}
\delta \bold{L}=\frac{\delta \bold{L}}{\delta \Phi}\delta\Phi+d\boldsymbol{\Theta}\left(\delta \Phi\right)
\end{equation}
where $\Phi$ denotes all fields (in our case just $g_{ab}$ and $A_{a}$), and specialising it to a fluctuation generated by $(\xi^{a},\mathcal{P})$ as in \eqref{gauge} one finds, on the one hand,
 \begin{equation}
 \delta_{\xi,\mathcal{P}}\bold{L}=-\pounds_{\xi}\bold{L}=-d\left(\iota_{\xi}\bold{L}\right).
 \end{equation}
On the other hand, from Noether's second theorem \cite{Barnich:2001jy},
 \begin{equation}
 \frac{\delta\bold{L}}{\delta\Phi}\delta_{\xi,\mathcal{P}}\Phi=d\bold{S}_{\xi,\mathcal{P}},
 \end{equation}
 where $\bold{S}_{\xi,\mathcal{P}}$ is a 3-form that is homogeneous in $\delta \bold{L}/\delta \Phi$ and its derivatives. It follows that
 \begin{equation}
 d\left(\boldsymbol{\Theta}\left(\delta_{\xi,\mathcal{P}}\Phi\right)+\iota_{\xi}\bold{L}+\bold{S}_{\xi,\mathcal{P}}\right)=0,
 \end{equation}
and this leads to the so-called Noether--Wald charge $\bold{Q}_{\xi,\mathcal{P}}$ associated to $(\xi^{a},\mathcal{P})$, defined by (existence of such $\bold{Q}_{\xi,\mathcal{P}}$, and that it is a local function of the fields, is guaranteed given some technical assumptions \cite{Homotopy} which are meet in our theory \eqref{lagrangian})
 \begin{equation}\label{NW}
 d\bold{Q}_{\xi,\mathcal{P}}=\boldsymbol{\Theta}\left(\delta_{\xi,\mathcal{P}}\Phi\right)+\iota_{\xi}\bold{L}+\bold{S}_{\xi,\mathcal{P}}.
 \end{equation}
These quantities can be computed for the Lagrangian \eqref{lagrangian}, and read
\begin{align}
16\pi\frac{\delta \bold{L}}{\delta \Phi}\delta\Phi&=\mathcal{E}_{ab}\delta g^{ab}\boldsymbol{\epsilon}+\bold{E}\wedge\delta A\\
16\pi\boldsymbol{\Theta}\left(\delta\Phi\right)&=-4\star F\wedge\delta A+16\pi \boldsymbol{\Theta}^{GR}\\
16\pi \bold{S}_{\xi,\mathcal{P}}&=\mathcal{P}\bold{E}+2\xi_{b}\mathcal{E}^{ab}\boldsymbol{\epsilon}_{a}\\
16\pi \bold{Q}_{\xi,\mathcal{P}}&=16\pi \bold{Q}^{GR}_{\xi}+4 \mathcal{P}\star F
\end{align}
where 
\begin{align}
\mathcal{E}_{ab}&=G_{ab}-T^{EM}_{ab}\\
\bold{E}&=4d\star F\\
T_{ab}^{EM}&=F_{ac}F_{b}^{\ c}+\star F_{ac}\star F_{b}^{\ c}\\
16\pi \boldsymbol{\Theta}^{GR}&=\star \theta^{GR}\\
\theta^{GR}_{a}&=\left(\delta^{c}_{a}\delta^{d}_{b}-g_{ab}g^{cd}\right)\delta \tensor{\Gamma}{^b _c_d}\\
16\pi\bold{Q}^{GR}_{\xi}&=\star d\xi
\end{align}
and the notation $\boldsymbol{\epsilon}_{a}T^{a}$ means contraction with the first index of the volume form, while for the Hodge dual we use $\star T_{abc}=\boldsymbol{\epsilon}_{abcd}T^{d}$. The fundamental identity follows from the first variation of \eqref{NW} given a fixed choice of parameters $\xi^{a}$ and $\mathcal{P}$, that is, $\delta\xi^{a}=0$ and $\delta P=0$ (one could be more general as in \cite{Ortin:2022uxa} and allow $\mathcal{P}$ to be field-dependent and have a non-trivial variation, see also \cite{Prabhu:2015vua}. However, the final results are unchanged and we find it more convenient to proceed as indicated). Writing the variation in the most convenient form, though, requires some massaging. Consider first the term $\iota_{\xi}\delta\bold{L}$. One has 
\begin{equation}
\iota_{\xi}\delta\bold{L}=\pounds_{\xi} \boldsymbol{\Theta}\left(\delta\Phi\right)-d\left(\iota_{\xi} \boldsymbol{\Theta}\left(\delta\Phi\right)\right)+\iota_{\xi}\left(\frac{\delta \bold{L}}{\delta\Phi}\delta\Phi\right)\, ,
\end{equation}
where we used \eqref{1stvar} and Cartan's formula $\pounds_{\xi}=\iota_{\xi}d+d\iota_{\xi}$. In turn, $\pounds_{\xi} \boldsymbol{\Theta}\left(\delta\Phi\right)$ can be written as
\begin{equation}
\begin{aligned}
\pounds_{\xi} \boldsymbol{\Theta}\left(\delta\Phi\right)&=-\delta_{\xi,\mathcal{P}} \boldsymbol{\Theta}\left(\delta\Phi\right)+\frac{1}{16\pi}\left(\iota_{\xi}\delta A\right)\bold{E}-\frac{1}{4\pi}d\left[ \left(\iota_{\xi}\delta A\right) \star F\right]
\end{aligned}
\end{equation}
where we have defined $\delta_{\xi,\mathcal{P}}\delta A$ according to 
\begin{equation}\label{varvar}
\delta_{\xi,\mathcal{P}}\delta A=\delta_{\xi,\mathcal{P}}(A'-A)=-(\iota_{\xi}F'+d\mathcal{P})+(\iota_{\xi}F+d\mathcal{P})=-\iota_{\xi}\delta F.
\end{equation}
Putting all these together, the first variation of \eqref{NW} can be cast in the form
\begin{equation}\label{mastid}
d\left(\delta\bold{Q}_{\xi}^{GR}+\iota_{\xi}\boldsymbol{\Theta}^{GR}+\frac{1}{4\pi}\mathcal{P}\delta\left(\star F\right)-\frac{1}{4\pi}\iota_{\xi}\left(\star F\right)\wedge \delta A\right)=\boldsymbol{\omega}\left(\delta \Phi,\delta_{\xi,\mathcal{P}}\Phi\right)+\iota_{\xi}\left(\frac{\delta \bold{L}}{\delta\Phi}\delta\Phi\right)+\frac{\iota_{\xi}\delta A}{16\pi}\bold{E}+\delta \bold{S}_{\xi,\mathcal{P}},
\end{equation}
where 
\begin{equation}\label{tal}
\boldsymbol{\omega}\left(\delta \Phi,\delta_{\xi,\mathcal{P}}\Phi\right)\equiv\delta\boldsymbol{\Theta}\left(\delta_{\xi,\mathcal{P}}\Phi\right)-\delta_{\xi,\mathcal{P}} \boldsymbol{\Theta}\left(\delta\Phi\right).
\end{equation}
It is important to notice that, from \eqref{gauge} and \eqref{varvar}, one has $[\delta,\delta_{\xi,\mathcal{P}}]=0$ and this implies that $\boldsymbol{\omega}\left(\delta \Phi,\delta_{\xi,\mathcal{P}}\Phi\right)$ is skew-symmetric and bilinear in $\delta\Phi$ and $\delta_{\xi,\mathcal{P}}\Phi$ (what can be verified from direct computation, too).

Consider now an asymptotically flat black hole solution of \eqref{lagrangian}, whose event horizon $\mathcal{H}$ is a Killing horizon of some vector field $k^{a}$, and assume that $k^{a}$ is a symmetry of the Maxwell field too, in the gauge-invariant sense $\pounds_{k}F=0$. Then, one has the associated electric and magnetic momentum maps $\mathcal{P}_{k}$ and $ \tilde{\mathcal{P}}_{k}$, defined as in \eqref{mm}, that satisfy a zeroth law $(\mathcal{P}_{k}\vert_{\mathcal{H}},\tilde{\mathcal{P}}_{k}\vert_{\mathcal{H}})=\text{constant}$ \cite{Ortin:2022uxa}. In such background, setting $(\xi^{a},\mathcal{P})=(k^{a},\mathcal{P}_{k})$ one has
\begin{equation}\label{sym}
\delta_{k}g_{ab}=0\, , \ \ \ \ \ \delta_{k,\mathcal{P}_{k}}A=0\, ,
\end{equation}
and we emphasise that no assumption has been made on the gauge of $A_{a}$. Thus, for a generic fluctuation \eqref{mastid} reduces to
\begin{equation}\label{bgonshell}
d\left(\delta\bold{Q}_{k}^{GR}+\iota_{k}\boldsymbol{\Theta}^{GR}+\frac{1}{4\pi}\mathcal{P}_{k}\delta\left(\star F\right)-\frac{1}{4\pi}\iota_{k}\left(\star F\right)\wedge \delta A\right)=\delta \bold{S}_{k,\mathcal{P}_{k}}.
\end{equation}
The cancellations on the right hand side follow from the fact that the background space is on-shell, so $\mathcal{E}_{ab}=0$ and $\bold{E}=0$, and because of \eqref{sym} and the fact that $\boldsymbol{\omega}\left(\delta \Phi,\delta_{k,\mathcal{P}_{k}}\Phi\right)$ is homogeneous in $\delta_{k,\mathcal{P}_{k}}\Phi$. Next, the idea is to evaluate the integral of \eqref{bgonshell} on some suitable 3-surface $\Sigma$, with boundaries at the horizon and at infinity. However, given that one also wants to consider variations of the magnetic charge of the black hole, it is inconsistent to assume that $\delta A$ is regular outside the black hole (indeed, if $\delta F=d\delta A$ where $\delta A$ is regular everywhere on a 2-sphere that encloses the hole, then by Stokes theorem the integral of $\delta F$ on that sphere must vanish and there is no variation of magnetic charge). This introduces complications in applying Stokes theorem on the left hand side of \eqref{bgonshell}: the surface $\Sigma$ and its boundaries should be chosen such that $\delta A$ is smooth there. Instead, one can rewrite \eqref{bgonshell} in a more convenient form by noticing that
\begin{equation}
-\iota_{k}\left(\star F\right)\wedge \delta A=d\tilde{\mathcal{P}}_{k}\wedge\delta A=-\tilde{\mathcal{P}}_{k}\wedge \delta F+d\left(\tilde{\mathcal{P}}_{k}\wedge\delta A\right)\, .
\end{equation}
Then, in terms of the notation introduced in Section \ref{M}, equation \eqref{bgonshell} becomes 
\begin{equation}\label{bgonshellduality2}
d\left(\delta\bold{Q}_{k}^{GR}+\iota_{k}\boldsymbol{\Theta}^{GR}+\frac{1}{4\pi}\mathcal{P}^{I}_{k}\Omega_{IJ}\delta F^{J}\right)=\delta \bold{S}_{k,\mathcal{P}_{k}}\, ,
\end{equation}
thus making manifest both gauge- and duality-invariance. The quantities $\delta F^{I}$ are physical (and therefore regular), so no issue arises in applying Stokes theorem on the left hand side of \eqref{bgonshellduality2}.

Consider now a fluctuation that is a solution of the (linearised) Einstein--Maxwell equations in the presence of a linear source that has both electric and magnetic charge,
\begin{equation}\label{pertnonfixed}
\begin{aligned}
\delta(G_{ab}-T^{EM}_{ab})&=8\pi T_{ab}\\
d\delta F^{I}&=-4\pi  \star j^{I}
\end{aligned}
\end{equation}
In principle, these equations contradict the hypotheses in which \eqref{bgonshellduality2} has been derived, since we assumed local existence of $\delta A$ (even though no assumption has been made about its properties), which is incompatible with $d\delta F\ne0$. This can be circumvented by making the sources purely electric via a duality rotation (e.g.~for a particle this would be achieved with a rotation of angle $\alpha=-\arctan(g/e)$), apply \eqref{bgonshellduality2} and then rotate the result back to a general duality frame.\footnote{An alternative is to repeat our derivation starting from a ``democratic'' formulation of electromagnetism \cite{deWit:2005ub,Meessen:2022hcg}, although the results should be precisely the same.} A simple computation yields that the right hand side of \eqref{bgonshellduality2} is
\begin{equation}
    \delta \bold{S}_{k,\mathcal{P}_{k}}=-\mathcal{P}_{k}^{I}\Omega_{IJ}\star j^{J}+k_{a}T^{ab}\boldsymbol{\epsilon}_{b},
\end{equation}
so one is left with
\begin{equation}\label{DIFMAST}
d\left(\delta\bold{Q}_{k}^{GR}+\iota_{k}\boldsymbol{\Theta}^{GR}+\frac{1}{4\pi}\mathcal{P}^{I}_{k}\Omega_{IJ}\delta F^{J}\right)=-\mathcal{P}_{k}^{I}\Omega_{IJ}\star j^{J}+k_{a}T^{ab}\boldsymbol{\epsilon}_{b}\, .
\end{equation}
The only thing remaining is to integrate \eqref{DIFMAST} on a suitable hypersurface $\Sigma$. We shall take it as a simply-connected 3-surface with boundary $\partial{\Sigma}=S^{2}_{\mathcal{H}}\sqcup S_{\infty}^{2}$, where $S_{\infty}^{2}$ is a 2-sphere at infinty and $S^{2}_{\mathcal{H}}$ is some spatial 2-sphere at the horizon. Using Stokes theorem on the left hand side, we have 
\begin{equation}
\begin{aligned}
\int_{S^{2}_{\infty}}\left[\delta\bold{Q}_{k}^{GR}+\iota_{k}\boldsymbol{\Theta}^{GR}+\frac{1}{4\pi}\mathcal{P}^{I}_{k}\Omega_{IJ}\delta F^{J}\right]=&\int_{S^{2}_{\mathcal{H}}}\left[\delta\bold{Q}_{k}^{GR}+\iota_{k}\boldsymbol{\Theta}^{GR}+\frac{1}{4\pi}\mathcal{P}^{I}_{k}\Omega_{IJ}\delta F^{J}\right]\\
&+\int_{\Sigma}\left[-\mathcal{P}_{k}^{I}\Omega_{IJ}\star j^{J}+k_{a}T^{ab}\boldsymbol{\epsilon}_{b}\right]\, .
\end{aligned}
\end{equation}
Assuming that the fluctuations $(\delta g_{ab},\delta F^{I})$ are regular and asymptotically flat, the integral at infinity gives
\begin{equation}
\int_{S^{2}_{\infty}}\left[\delta\bold{Q}_{k}^{GR}+\iota_{k}\boldsymbol{\Theta}^{GR}+\frac{1}{4\pi}\mathcal{P}^{I}_{k}\Omega_{IJ}\delta F^{J}\right]=\int_{S^{2}_{\infty}}\left[\delta\bold{Q}_{k}^{GR}+\iota_{k}\boldsymbol{\Theta}^{GR}\right]=\Omega_{H}\delta J-\delta M
\end{equation}
where in the first step we used that $\mathcal{P}^{I}_{k}$ satisfy the asymptotic boundary condition \eqref{boundaryCond}, and the last equality was established in \cite{Iyer:1994ys} (see also \cite{Barnich:2001jy}), where $M$ and $J$ are the ADM mass and angular momentum. At the horizon, we have
\begin{equation}
\int_{S^{2}_{\mathcal{H}}}\left[\delta\bold{Q}_{k}^{GR}+\iota_{k}\boldsymbol{\Theta}^{GR}+\frac{1}{4\pi}\mathcal{P}^{I}_{k}\Omega_{IJ}\delta F^{J}\right]=-\phi^{I}\Omega_{IJ}\delta Q^{J}+\int_{S^{2}_{\mathcal{H}}}\left[\delta\bold{Q}_{k}^{GR}+\iota_{k}\boldsymbol{\Theta}^{GR}\right]
\end{equation}
where we used the constancy of $\mathcal{P}^{I}_{k}$ at $\mathcal{H}$ by virtue of the zeroth law, and have introduced the electromagnetic potentials $\phi^{I}=-\mathcal{P}^{I}_{k}\vert_{\mathcal{H}}$. We thus get the desired identity
\begin{equation}\label{INTMAST}
\delta M-\Omega_{H}\delta J=\phi^{I}\Omega_{IJ}\delta Q^{J}-\int_{S^{2}_{\mathcal{H}}}\left[\delta\bold{Q}_{k}^{GR}+\iota_{k}\boldsymbol{\Theta}^{GR}\right]+\int_{\Sigma}\left[\mathcal{P}^{I}_{k}\Omega_{IJ}\star j^{J}-k_{a}T^{ab}\boldsymbol{\epsilon}_{b}\right].
\end{equation}

\bibliography{ref}

\end{document}